%% file: conference_101719.tex
\documentclass[conference]{IEEEtran}
\IEEEoverridecommandlockouts
\usepackage{cite}
\usepackage{amsmath,amssymb,amsfonts}
\usepackage{algorithmic}
\usepackage{graphicx}
\usepackage{textcomp}
\usepackage{xcolor}
\usepackage{multirow}
\usepackage{subcaption}

\usepackage{todonotes}
\def\BibTeX{{\rm B\kern-.05em{\sc i\kern-.025em b}\kern-.08em
    T\kern-.1667em\lower.7ex\hbox{E}\kern-.125emX}}
\usepackage[singlespacing]{setspace} 

\def\varM{M}

\begin{document}

\title{Semantically Optimized End-to-End Learning for Positional Telemetry in Vehicular Scenarios\\
}
\author{\IEEEauthorblockN{Neelabhro Roy, Samie Mostafavi \& James Gross}
\IEEEauthorblockA{\text{School of Electrical Engineering and Computer Science} \\
\text{KTH Royal Institute of Technology}\\
Stockholm, Sweden \\
\{nroy, ssmos, jamesgr\}@kth.se}
}

\maketitle

\input{abstract.tex}
\input{introduction.tex}

\input{Related}

\input{sys_model}

\input{approach}

\input{results}

\input{conclusions}

\bibliographystyle{IEEEtran}
\bibliography{IEEEabrv,references}

\end{document}

%% file: abstract.tex
\begin{abstract}
End-to-end learning for wireless communications has recently attracted much interest in the community, owing to the emergence of deep learning-based architectures for the physical layer. Neural network-based autoencoders have been proposed as potential replacements of traditional model-based transmitter and receiver structures. Such a replacement primarily provides an unprecedented level of flexibility, allowing to tune such emerging physical layer network stacks in many different directions. The semantic relevance of the transmitted messages is one of those directions. In this paper, we leverage a specific semantic relationship between the occurrence of a message (the source), and the channel statistics. Such a scenario could be illustrated for instance, in vehicular communications where the distance is to be conveyed between a leader and a follower. We study two autoencoder approaches where these special circumstances are exploited. We then evaluate our autoencoders, showing through the simulations that the semantic optimization can achieve significant improvements in the BLERs (up till 93.6\%) and RMSEs (up till 87.3\%)  for vehicular communications leading to considerably reduced risks and needs for message re-transmissions.
\end{abstract}

%% file: introduction.tex
\section{Introduction}
\label{sec:intro}
Softwarization of the wireless network stack has for a long time been an important research goal pursued by the community. 
On the research side, this materialized over a decade ago with the introduction of Software-Defined Radios, which despite their different architectures represented a big step forward in introducing more flexible applications by leveraging software-based implementations. Since then, the quest for more and more flexible wireless transceiver structures has steadily evolved and has lately been strongly influenced by machine learning.
Especially the significant breakthroughs in neural networks for classification and pattern recognition associated with the advances in training acceleration through the use of GPUs about ten years ago, have led to an enormous interest in using machine learning for various problems in communication systems.
In particular, deep learning (DL) has recently shown great potential to become a powerful tool to design, optimize, adapt, and secure wireless communications and in doing so, introducing an unprecedented level of flexibility in the transceiver structures.

Machine learning applications to communication systems' design have thus received significant research attention recently. 
With respect to the implemented architectures, works either focus on substituting individual functions in the transceiver chains, or more progressively substituting larger blocks, primarily in the physical layer.
Examples of the first category comprise for instance works on signal detection~\cite{Samuel_2019}, channel estimation~\cite{Neumann_2018}, or signal demapping in broadband wireless communication systems~\cite{Shental_2019}.
In all cases, it can be shown that deep learning, given sufficient training data, is either at par with legacy (model-based) approaches, or even outperforms them.
Depending on the application, this can come with a lower complexity of the learning approach. 

In contrast to learning individual transceiver functions, substituting larger functions can achieve more flexibility~\cite{Honkala_2021}.
The most flexible approach to date is to substitute entire blocks of transceivers by so called end-to-end approaches~\cite{dorner_deep_2018}.
Here, in contrast to~\cite{Honkala_2021}, the entire transmitter and receiver are substituted through DNNs, which allows for highly flexible signalling schemes.
In detail, variational autoencoders are utilized for the joint training of transmitter and receiver.
Such end-to-end approaches are most consequent in moving away from model-based transceivers, potentially jeopardising traditional system standardization.

End-to-end data-driven communication methods provide principally new ways to refine communication systems due to the high degree of flexibility that they introduce. 
Among other directions, semantic optimization of end-to-end learning systems is a promising one. 
This idea goes back to Shannon and Weaver \cite{shannon} in which the primary focus is how precisely transmitted symbols over an erroneous channel convey some desired meaning. 
While classical information theory concerns only successful transmission of symbols from transmitter to receiver and improving bit error rate (BER), semantic communication systems are tuned towards minimizing the semantic error instead of BER.
Leveraging these ideas in the context of end-to-end learning for communication systems has been recently addressed by several works ~\cite{semantic}.
Authors in ~\cite{xie_deep_2021} and ~\cite{zhou_semantic_2022} devised a semantic communication scheme for the text transmission scenario over channels with various signal-to-noise ratios (SNRs).
Utilizing natural language processing (NLP) techniques combined with DL, their system essentially encodes and decodes the meaning of each input sentence similar to a source codec.
In ~\cite{xuan_deep_2021}, authors consider the scenario of correlated sources and devise a deep joint source-channel codec for AWGN channels.
They model the source by the Gauss-Markov process and the temporal information is extracted by utilizing recurrent neural networks (RNNs).

In contrast to these works, in this paper we assume a different scenario where there is a semantic relation between the source message and the channel statistics.
Such a relation can be illustrated in positional telemetry or position-related transmission scenarios where there is a relation between the transmitted information and the channel quality.
For example, in Vehicle-to-vehicle (V2V) communications, it becomes all the more important to ensure correct reception of messages between the vehicles, as they get closer to each other. To exemplify our proposed approach, we consider a V2V communication based set-up with a leader-follower scenario where the distance between them is estimated by the follower and subsequently transmitted to the leader. 
In this set-up, the transmitted message directly correlates to the channel statistic, since for higher distances the SNR is generally lower. We show that in such a set-up a large optimization potential results from building standard end-to-end autoencoders from training samples that exhibit a relationship between the message semantic and the channel state.   
This advantage persists over a substantial parameter range, and can be further tuned by modifications of the loss function. 

The rest of the paper is structured in the following manner. 
In Section~\ref{sec:sys_model} we first introduce the detailed system model, before we discuss in Section~\ref{sec:approach} our main approach.
Numerical results are presented in Section~\ref{sec:results}, while the paper is concluded in Section~\ref{sec:conclusions}.


%% file: Related.tex

%% file: sys_model.tex
\section{System Model}
\label{sec:sys_model}
Our work considers a simple set-up with a single-antenna transmitter, a channel and a single-antenna receiver as described in Fig. \ref{arch}.
The transmitter communicates a message: \[ s \hspace{2mm}  \epsilon \hspace{2mm}  \varM = {1,2, .... M}\] to the receiver, across the channel. 
To realize this, $n$ complex baseband symbols are transmitted forming a vector $\mathbf{x} \in \mathbb{C}^n$ under a power constraint.
Each message $s$ can be represented by a sequence of bits of
length $k = log_2(M)$. Hence, the resulting communication rate can be obtained as $R = k/n$ in bits/channel use.
The channel $h$ between transmitter and receiver introduces path loss and shadow fading effects, leading to a substantial distortion of the originally sent message.
Path loss and shadow fading effects are assumed to impact all baseband symbols of a sent message equally, but vary from message to message.
At the receiver, additive white Gaussian noise $\mathfrak{n}$ further corrupts the transmitted symbols, such that the receiver ends up with the received vector $\mathbf{y} \in \mathbb{C}^n$.
The resulting signal model is thus given as:
\[ \mathbf{y} = h \cdot \mathbf{x} + \mathfrak{n}. \]
From this receive statistic, an estimate $\hat{s}$ of $s$ is obtained, where the corresponding block-error rate (BLER) is obtained as:
\[ P_e = 1/M* \sum_s Pr(\hat{s} \neq s | s).\]
Assuming the transmit power to be denoted by $P_{tx}$ while the noise variance is given by $\sigma^2$, the SNR of a received message results to \[ \gamma = P_{tx} * h^2/\sigma^2.\]
In this work we consider circumstances that map the occurrence of a message $s$ to a corresponding channel attenuation $h$. 
As an example scenario consider a leader-follower set-up in a vehicular environment, where the follower may wish to convey its distance $d$ towards the leader through a wireless system, as shown in Fig. \ref{arch}.
Given a certain distance estimate, the approximate channel gain $h$ is specific to the distance estimate.  Likewise, for a given attenuation $h$ due to path loss and shadow fading only a certain set of messages occur.
The scenario can be extended towards the communication of position estimates or positional telemetry to fixed infrastructure nodes (V2I communication), or other communication occurrences where for a spatial region only subsets of the message set might occur.
In the chosen set-up, beyond the BLER as the standard performance metric, due to the semantic of the distance between vehicles, the root mean squared error between the transmitted and estimated message is relevant. 
Our interest is thus in finding complex symbol representations that exploit the specific channel-message relationship while the semantic error between sent and estimated message is more relevant than a pure BLER metric. This motivates us to come up with a loss function incorporating the RMSEs between the messages. Additionally, we choose the Root Mean Squared Error (RMSE) as the loss metric to analyse the performance of our proposed model
The RMSE can be defined as:
\[ P_{RMSE} = \sqrt{ \sum_s (\hat{s} - s)^2/N} \]


%% file: approach.tex
\section{Architecture and Approach}
\label{sec:approach}

In this section we present our system architecture and adopted approaches to realize it.

\subsection{Baseline Approach}
\label{subsec:baseline}
Dörner \textit{et al.} in \cite{dorner_deep_2018} proposed an end-to-end learning based communication system, built, trained and run solely on deep neural networks, using unsynchronized off-the-shelf software defined radios (SDRs), which acts as our baseline approach. 
The approach aims to reproduce the input at the transmitter as the output at the receiver end, reinforcing the underlying idea of end-to-end learning. 
The transmitter in their proposed autoencoder architecture comprises of an embedding layer with an exponential linear unit (ELU) activation succeeded by a feedforward neural network layer whose 2$n$-dimensional output is cast to an $n$ dimensional complex-valued vector by considering real and imaginary halves.
A normalization layer forms the last layer of the transmitter ensuring the power constraints on  the encoder output $x$ are met.
For the channel characteristics, an additive white Gaussian noise (AWGN) channel trained at a constant Signal-to-Noise-Ratio(SNR) is chosen.
Their proposed receiver concatenates the real and imaginary parts of the preceding channel output to a real output, which is succeeded by a 'softmax' activated feedforward neural network. The output of this layer:
\[ b \hspace{2mm}  \epsilon \hspace{2mm}  (0,1)^M \] is a probability vector having corresponding probabilities of all the possible messages. The final estimate $\hat{s}$ of $s$ is derived from the largest index amongst the elements of $b$. The autoencoder is consequently trained using stochastic gradient descent (SGD). They use the cross-entropy loss function as the task at hand is of classification. The resulting loss function is thus:
\[ L_{Loss-Baseline} = -log(b_i) \] where $b_i$ corresponds to the $i^{th}$ element of the vector $b$.

\begin{figure}[!t]
\centering
\includegraphics[width =3.4in, height=1.4in]{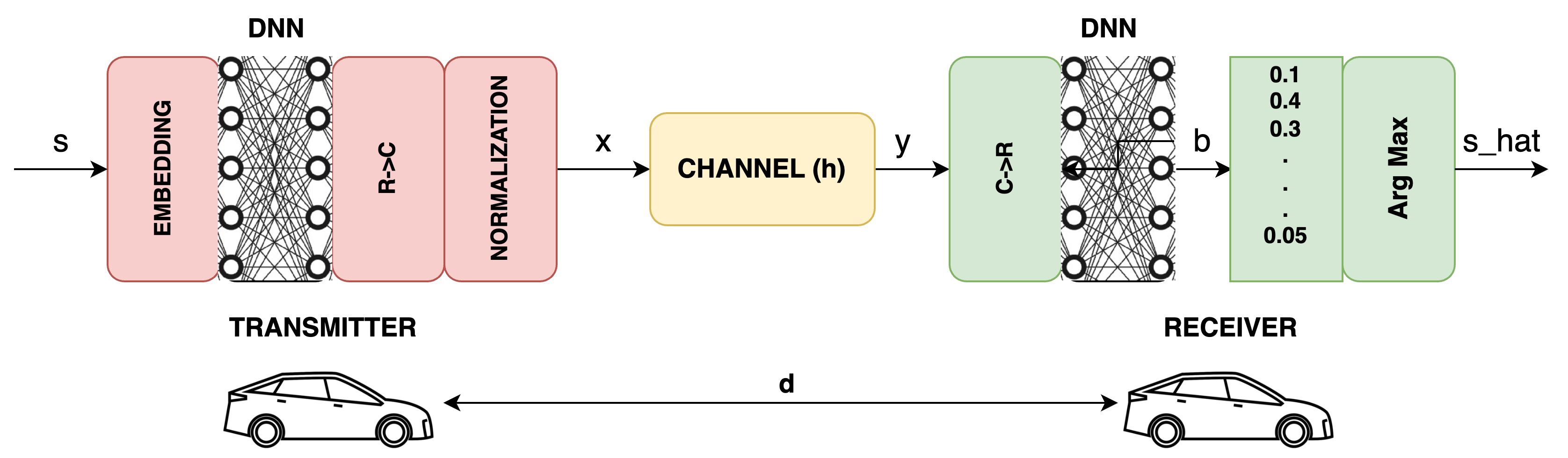}
\caption{Autoencoder architecture}
\label{arch}
\end{figure}

\subsection{Semantically Optimized System}
\label{subsec:unweighted}
The idea behind our first proposed system is to study the impact for cases where messages occur only in conjunction with certain channel statistics, as described in Section~\ref{sec:sys_model} for transmitting the distance in a leader-follower scenario. 
For a given distance, the transmission is distorted by the distance-dependent path loss as well as a random shadow fading component (in addition to the noise at the receiver). 
The resulting receive statistic is specific to the distance.
Assuming $M$ to encode a certain number of distance settings, we obtain a semantically optimized system by training the autoencoder from Section~\ref{subsec:baseline} with the specific channel for message $s$. 

Fig.~\ref{arch} exemplifies this semantically optimized architecture, where the input message sequence $s$ is fed into the transmitter, leading to an output of $x$ from the transmitter. The output vector then passes through the channel $h$, which is dependant on the input $s$. 
The vector $h$ creates a mapping of each message comprising $s$, to the corresponding channel conditions emanating out of the path loss model of our choice. 
The output of the channel $y$ is then fed to the receiver where after passing through the DNNs and subsequent softmax activation, we obtain our estimate $\hat{s}$ of the original input signal $s$.

\begin{figure}[!t]
\centering
\includegraphics[width =3.6in, height=0.8in]{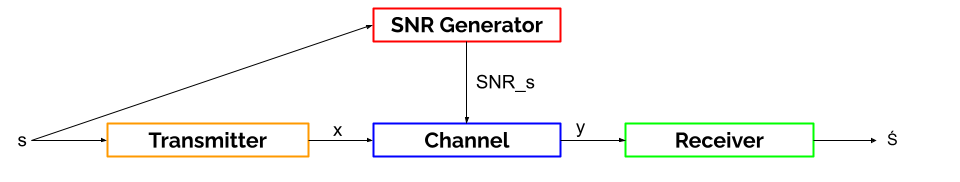}
\caption{Semantically Optimized Autoencoder simulation scenario}
\label{SPL_arch}
\end{figure}

\subsection{Weighted-semantically Optimized System}
\label{subsec:weighted}
As described before, the neural network architecture is designed to minimize the cross-entropy loss incurred during the encoding and decoding process. Our weighted-semantically optimized system builds on top of the aforementioned semantically optimized system, by bringing in an additional layer of semantic awareness when computing. 

We propose a combination of the cross entropy loss function and a modified root-mean-squared-loss function as our loss function, incorporating the labels $s$. This is used since the end-to-end learning communication problem has been framed as a classification task (and hence, the use of cross-entropy loss), but at the same, it also aims to minimize the loss between the end-to-end metrics. The resulting function is thus:
\[ L_{Loss-SPL} = -log(b_i) - 1/s * \sqrt{ \sum_{i=0}^{M} {b_i}(s-i)^2} \] where $b_i$ corresponds to the $i^{th}$ element of the vector $b$.

We will explore how introducing the newly proposed loss function while computing the cross entropy loss in combination with our modified RMSE loss function pays dividends in terms of the associated RMSEs when compared to the semantically optimized approach and the baseline approach.

%% file: results.tex
\section{Results}
\label{sec:results}

In this section, we present our simulation results evaluating our proposed model by analysing different loss metrics such as the Root Mean Squared Errors (RMSEs), the Block Error Rates (BLERs) and the associated signal constellations in various experimental settings.
\subsection{Methodology}
In this subsection, we describe our experimental methodology and evaluate these approaches:
\begin{itemize}
  \item For our baseline implementation we use the end-to-end learning model as proposed in \cite{dorner_deep_2018} incorporating additive white Gaussian Noise (AWGN) at a constant SNR of 7dB. 
  \item We then proceed to evaluate the proposed semantic path loss model trained at adaptive SNRs, varying according to the distance between the transmitter and receiver as mentioned in Section~\ref{subsec:unweighted}. This scheme is referred to as 'SPL' in the forthcoming plots.
  \item Finally, we evaluate the proposed weighted semantic path loss model, building on top of the semantic pathloss model by weighting the transmitted sequence in the manner as described in Section~\ref{subsec:weighted}. We refer to this approach as 'Weighted SPL' in the plots.
\end{itemize}
We utilize Python's open source libraries such as Tensorflow \cite{DBLP:journals/corr/AbadiBCCDDDGIIK16} and NumPy  \cite{numpy} to realize our approach and our simulation scenario is shown in Fig. \ref{SPL_arch}. 
For the pathloss, we utilize the standard model
\[ 10log_{10} * \lambda /(4\pi d)^{\phi}\]
setting $\lambda = 0.05cm$ in the formula mentioned for SNR, for an IEEE 802.11p scenario with $\phi = 2.8$
We implement log-normal shadowing model in the channel with a standard deviation of 3dB.  Thus, our results show a comparison of our proposed models and other legacy approaches over varying channel conditions.

To train the baseline approach and our proposed model, we employ multi batch training, encompassing 3 batches of 200,000 samples each, with learning rates for each batch successively decreasing from 0.1 to 0.01 and finally to 0.001. Each batch also iterates 10,000 times while training. For the validation set, we use similar settings. We evaluate these models with a random uniform sequence of messages $s$, where:
\[ s \hspace{2mm}  \epsilon \hspace{2mm} {1, 2, 3,  .... 256} \]
corresponding to $k = 8$. The length of the transmitted message sequence equals the batch size of the experimental run.
We begin by studying the transmission of sequences where one message is composed of $n = 2$ complex symbols with $M = 256$ as also evidenced by $k = 8 (k = log_2(M))$. These parameters are summarized in Table \ref{Tab:param}.

\begin{table}[h!]
\begin{center}
\caption{Experimental Parameters}
\begin{tabular}{| p{0.5cm} | p{4.9cm} | p{1.3cm}  |} \hline
\textbf{S.No.} & \textbf{Parameters} & \textbf{Values}\\ \hline
1 & Batch Size & 200,000\\ \hline
2 & $M$ & 256\\ \hline
3 & $k$ & 8 \\ \hline
4 & Learning rate & 0.01 \\ \hline
5 & Iterations/Batch & 10,000  \\ \hline
6 & AWGN SNR for Baseline & 7dB \\ \hline
7 & Log Normal Shadowing standard deviation & 3dB\\ \hline
8 & Frequency of transmission & 5.9GHz\\ \hline
9 & Wavelength of transmission & 5cm\\ \hline
10 & Pathloss exponent & 2.8\\ \hline
11 & Noise Figure & -95dBm\\ \hline
12 & Transmitted power & 20-23dBm\\ \hline
\end{tabular}
\label{Tab:param}
\end{center}
\end{table}

To understand the impact of the transmit power and also the number of complex channel uses a message sequence can utilize, we vary them too.
We summarize the various testing scenarios in Table \ref{Tab:scheme}:
\begin{table}[h!]
\begin{center}
\caption{Evaluation schemes}
\begin{tabular}{ |p{2.4cm}|p{1.3cm}|p{1.3cm}| p{1.3cm}| }

 \hline
\textbf{Parameters} & \textbf{Scenario 1} & \textbf{Scenario 2} & \textbf{Scenario 3}\\
\hline
 Tx Power   & 100mW    & 200mW & 100mW\\
 \hline
 Complex symbols $n$  & 2  & 2 & 4\\
 \hline
\end{tabular}
\label{Tab:scheme}
\end{center}
\end{table}


\subsection{Discussion}
\begin{figure*}[htp]
  \begin{subfigure}{0.33\linewidth}
    \centering
    \includegraphics[width=1\linewidth]{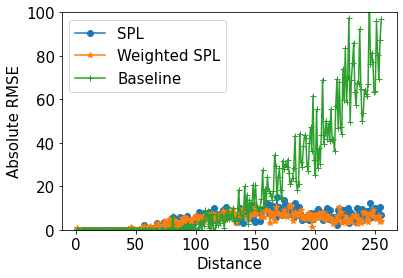}
    \caption{RMSEs under evaluation scenario 1}
    \label{fig:rmse_a}
  \end{subfigure}
  \begin{subfigure}{0.33\linewidth}
    \centering
    \includegraphics[width=1\linewidth]{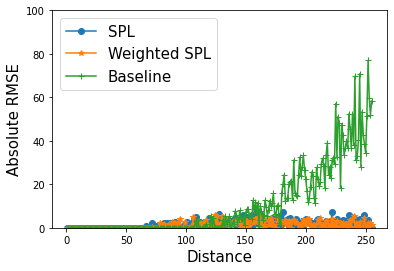}
    \caption{RMSEs under evaluation scenario 2}
    \label{fig:rmse_b}
  \end{subfigure}
  \begin{subfigure}{0.33\linewidth}
    \centering
    \includegraphics[width=1\linewidth]{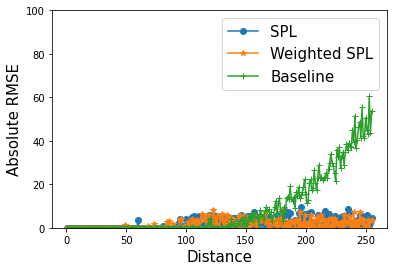}
    \caption{RMSEs under evaluation scenario 3}
    \label{fig:rmse_c}
  \end{subfigure}
  \caption{RMSEs under various evaluation scenarios}
  \label{fig:rmse}
\end{figure*}

The RMSEs for Scenario 1, Scenario 2 and Scenario 3 are shown in Fig. \ref{fig:rmse_a}, \ref{fig:rmse_b} and \ref{fig:rmse_c} respectively which individually compare the RMSEs for the baseline approach, our proposed semantic pathloss model and the proposed weighted semantic pathloss model. These plots are thus indicative of the variation of RMSEs with varying distances between the point of transmission and reception. Across all the subplots we can observe that the RMSEs in the baseline approach increase after a certain distance and maintain that trend, while for the semantic pathloss model and the proposed weighted semantic pathloss model the RMSEs are significantly less.


\begin{figure*}[htp]
  \begin{subfigure}{0.33\linewidth}
    \centering
    \includegraphics[width=1\linewidth]{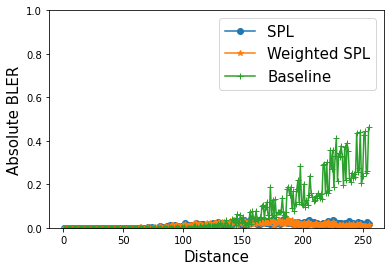}
    \caption{BLERs under evaluation scenario 1}
    \label{fig:bler_a}
  \end{subfigure}
  \begin{subfigure}{0.33\linewidth}
    \centering
    \includegraphics[width=1\linewidth]{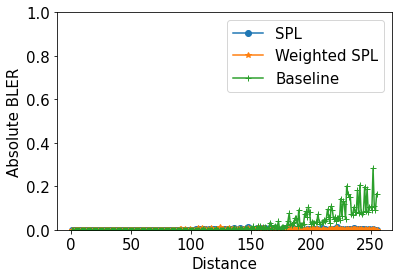}
    \caption{BLERs under evaluation scenario 2}
    \label{fig:bler_b}
  \end{subfigure}
  \begin{subfigure}{0.33\linewidth}
    \centering
    \includegraphics[width=1\linewidth]{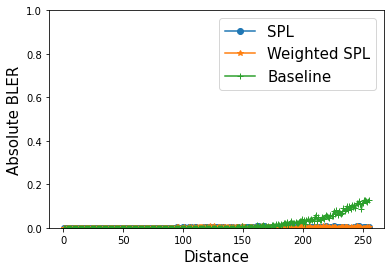}
    \caption{BLERs under evaluation scenario 3}
    \label{fig:bler_c}
  \end{subfigure}
  \caption{BLERs under various evaluation scenarios}
  \label{fig:bler}
\end{figure*}

Fig. \ref{fig:bler_a}, \ref{fig:bler_b} \& \ref{fig:bler_c} similarly show the associated BLERs under evaluation Scenario 1, Scenario 2 and Scenario 3 respectively, comparing the associated BLERs for the baseline approach, our proposed semantic pathloss model and the proposed weighted semantic pathloss model. These plots thus showcase the variation of the BLERs with varying distances between the transmitter and the receiver. Similar to the RMSE plots, in all the subplots we can observe that the BLERs in the baseline approach increase after a certain distance and maintain that trend, while for the semantic pathloss model and the proposed weighted semantic pathloss model the BLERs are much less.

\begin{figure}[htp]
\begin{center}
  \begin{subfigure}{0.49\linewidth}
    \centering
    \includegraphics[width=1\linewidth]{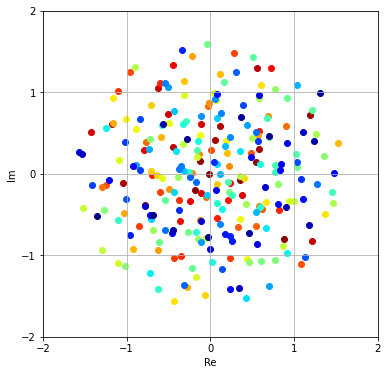}
    \caption{Baseline approach}
    \label{fig:const_a}
  \end{subfigure}
  \begin{subfigure}{0.49\linewidth}
    \centering
    \includegraphics[width=1\linewidth]{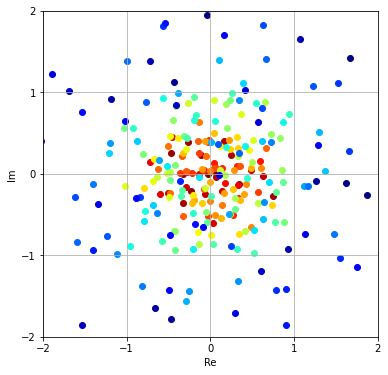}
    \caption{Semantic Pathloss}
    \label{fig:const_b}
  \end{subfigure}
  \begin{subfigure}{0.49\linewidth}
    \includegraphics[width=1\linewidth]{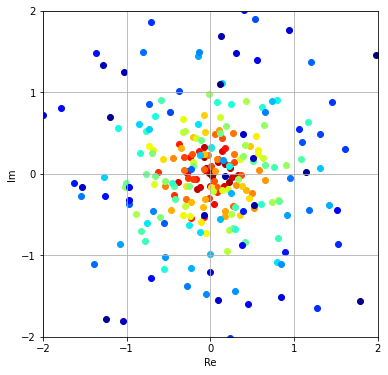}
    \caption{Weighted Semantic Pathloss}
    \label{fig:const_c}
  \end{subfigure}
  \caption{Signal constellations for various cases}
  \label{fig:const_100mW}
\end{center}  
\end{figure}

Fig. \ref{fig:const_a}, \ref{fig:const_b} and \ref{fig:const_c} showcase the scatter plot of the signal constellations for the baseline model, the semantic pathloss model and the weighted semantic pathloss model respectively, under evaluation scenario 1. Though there are two complex symbols in this scenario, we present the constellations for one of them, as both of the constellations for each case are largely similar.
We choose a spectral gradient to represent the messages in the constellations, to help understand the placement of the symbols by the autoencoder. As we move from the blue part of the spectrum (following VIBGYOR-Violet, Indigo, Blue, Green, Yellow, Orange and Red) to the red part, the messages $s$ these constellation points represent, decrease in terms of the distance. This means that the points in the blue part of the spectrum represent higher distances, and the points in the red part represent lower distances. 

Table \ref{Tab:rmse_Tab} provides the average RMSEs across all distances, under various testing schemes and also provides the percentage improvements of our schemes over the baseline approach. Similarly Table \ref{Tab:bler_Tab} provides the average BLER values over all distances when tested with our evaluation schemes.

\begin{table}[h!]
\begin{center}
\caption{Average RMSE summary}
\begin{tabular}{| p{1.5cm}| p{1.5cm} | p{1.5cm} | p{1.5cm}  |} \hline 

\textbf{Evaluation Scenario} & \textbf{Baseline Approach} & \textbf{SPL} & \textbf{Weighted SPL}\\ \hline

Scenario 1 & 20.04 & 4.79 (76\%) & 4.26 (78.7\%) \\ \hline

Scenario 2 & 11.42 & 1.54 (86.5\%) & 1.45 (87.3\%)\\ \hline

Scenario 3  & 10.63 & 1.68 (84.1\%) & 1.51 (85.7\%) \\ \hline
\end{tabular}
\label{Tab:rmse_Tab}
\end{center}
\end{table}

\begin{table}[h!]
\begin{center}
\caption{Average BLER summary}
\begin{tabular}{| p{1.5cm}| p{1.5cm} | p{1.8cm} | p{1.8cm}  |} \hline 

\textbf{Evaluation Scenario} & \textbf{Baseline Approach} & \textbf{SPL} & \textbf{Weighted SPL}\\ \hline 

Scenario 1 & 0.07  & 0.0129 (81.5\%) & 0.016 (77.1\%) \\ \hline 

Scenario 2 & 0.0282 & 0.0018 (93.6\%) & 0.0023 (91.8\%)\\ \hline 

Scenario 3 & 0.0239 & 0.0015 (93.7\%) & 0.0016 (93.3\%) \\ \hline
\end{tabular}
\label{Tab:bler_Tab}
\end{center}
\end{table}


As we can observe in Fig. \ref{fig:rmse_a}, our proposed autoencoders offer a significant improvement in the RMSEs over the baseline approach, with the weighted semantic path loss model slightly outperforming the semantic path loss model, in terms of the overall RMSEs incurred across the distances. As we move from Fig. \ref{fig:rmse_a} to \ref{fig:rmse_b}, we observe a reduction in the RMSEs which can be attributed to the increase in transmit power, allowing for a stronger signal to be transmitted, consequently leading to the availability of higher SNRs.
On comparing Fig. \ref{fig:rmse_a}  and \ref{fig:rmse_c} we again observe a drop in the RMSEs incurred, especially in the baseline approach largely due to the availability of more complex symbols to represent the messages $s$ spanning $M$, as $n$ increases from 2 to 4 across these plots. 
Hence, it can be concluded that increasing the transmit power offers a significant advantage in terms of higher SNR availability, consequently leading to lower RMSEs. Similar advantages can be obtained by allowing for more complex symbols to represent a transmitted sequence over our approach with 100mW transmit power and $n = 2$. The RMSEs in Fig. \ref{fig:rmse_b} are still more advantageous over those shown in Fig. \ref{fig:rmse_c} for our proposed approaches, if their summation across all distances are to be compared. From a practical standpoint, this drop in RMSEs leads to a much lower number of re-transmissions in order to ensure the correct reception of the messages, which we believe is a significant advantage of our proposed model over the baseline approach.


Across Fig. \ref{fig:bler_a}, \ref{fig:bler_b} and \ref{fig:bler_c} we can observe that our proposed models outperform the baseline approach by a significant margin.
On comparing Fig. \ref{fig:bler_a} and \ref{fig:bler_c} we observe a drop in BLERs across all models being tested, which can be attributed to the availability of more channel uses for message transmission. Between Fig. \ref{fig:bler_a} and \ref{fig:bler_b} as well, we can observe a drop because of the availability of higher SNRs owing to the increase in the transmit power.

Between the Weighted SPL approach and the SPL approach, we can observe a tradeoff between dropping RMSEs and increasing BLERs, with the introduction of our proposed loss function. This happens since our proposed loss function incorporates a modified RMSE in the loss function itself leading to lower RMSEs consistently, but increased BLERs.

In Fig. \ref{fig:const_a}, the constellations for the baseline approach are shown. We observe a generous intermixing of symbols belonging to both higher and lower distance messages $s$ spanning $M$, indicative of the largely equal importance the autoencoder provides to the symbols, to reduce the average BLER irrespective of their magnitudes. 

Moving on from Fig. \ref{fig:const_a} to Fig. \ref{fig:const_b}, we observe rather segregated signal constellations as the autencoder is now semantically empowered and places the messages indicative of lower distances(and having comparatively higher SNRs) closely towards the centre in a somewhat circular manner, and the messages indicative of longer distances mostly outside the aforementioned circle.
On moving from Fig. \ref{fig:const_b} to Fig. \ref{fig:const_c}, a contraction in the radii of the set of constellation points indicative of smaller messages can be observed, as a result, the points representing smaller distances are more concentrated towards the centre because of the introduction of the weighted loss function


%% file: conclusions.tex
\section{Conclusions \& Future Work}\label{sec:conclusions}
The aim of this work was to examine end-to-end data-driven communication methods when the semantic knowledge of the channel conditions can be utilized. It can be motivated that in a V2V communication scenario, it is critical that the communication between the vehicles or potentially infrastructural components remains error and risk free, as they get closer, in order to prevent accidents.
Thus, through the course of the work, a distance transmission scenario has been considered for a vehicular communications usecase, where the message directly correlates to the channel statistic, as for higher distances the SNR is generally lower.
Compared to state-of-the-art learning-based communication systems that are trained over constant channel conditions, we identified a large optimization potential to reduce RMSEs by only considering training samples where the message semantics correlates with the channel statistic.
Moreover, by proposing a new loss function in addition to already showing gains with the cross entropy loss function (up till 86.5\%), we observe a further improvement in the corresponding RMSEs (up till 87.3\%). Considering a practical scenario, this drop in RMSEs leads to a much lower number of re-transmissions needed in order to ensure the correct reception of the transmitted messages, where we believe our proposed model is significantly advantageous over the baseline approach.
Future works include utilizing this approach for positional telemetry amongst vehicles and towards fixed infrastructures and a study of various loss functions on the performance of such systems. The impacts of Doppler effect could be considered as an additional semantic information to extend the proposed methods.